\documentclass{article}

\usepackage{paralist}
\usepackage{graphicx}
\usepackage{fancyhdr}
\usepackage{times}
\usepackage{amssymb,amsmath}
\usepackage{color}
\usepackage{amsfonts}
\usepackage{dsfont}
\usepackage[latin1]{inputenc}
\usepackage{graphics}
\usepackage{latexsym,graphicx}

\title{Prediction of the margin of victory only from team rankings for regular season games in NCAA men's basketball}

\author{David Beaudoin\\
{\it david.beaudoin@fsa.ulaval.ca}\\
Département d'opérations et systèmes de décision\\
Université Laval\\
{ }\\
Thierry Duchesne\\
{\it thierry.duchesne@mat.ulaval.ca}\\
Département de mathématiques et de statistique\\
Université Laval}

\date{}
\IfFileExists{upquote.sty}{\usepackage{upquote}}{}
\begin{document}

\maketitle

\section*{Abstract}

The main objective of this paper is to investigate the extent to which the margin of victory can be predicted solely by the rankings of the opposing teams
in NCAA Division I men's basketball games. Several past studies have modeled this relationship for the games played during the March Madness tournament, and this work aims at verifying if the models
advocated in these papers still perform well for regular season games. Indeed,
most previous articles have shown that a simple quadratic regression model provides fairly accurate predictions of the margin of victory
when team rankings only range from 1 to 16. Does that still hold true when team rankings can go as high as 351? Do the model assumptions hold? Can we find semi- or non-parametric methods that yield even better results (i.e. predicted margins of victory that more closely resemble actual results)? The analyses presented in this paper suggest that the answer is ``yes'' on all three counts!

\bigskip

\noindent {\bf keywords}: generalized additive model; loess; nonparametric regression; quadratic regression; spatial smoothing

\section{Introduction}

Predicting the outcome of sporting events from team/player rankings is very popular, especially in tennis and NCAA Division I men's basketball. Most papers on this topic focus on who wins the match (i.e., they estimate winning probabilities); few consider the margin of victory as the response variable.
In tennis, del Corral \& Prieto-Rodriguez (2010) developed three probit models. The most significant explanatory variable turned out to be the difference in ATP/WTA rankings between the two opposing players. As a matter of fact, this was the only variable that ended up significant across all of their models. They also found that rank differences are more important among the top players for both men and women. In other words, the predicted winning probability increases in a more dramatic way between players ranked \#1 relative to \#11, compared to players ranked \#51 relative to \#61, for example.

In NCAA basketball, not only do most articles focus on binary win/loss indicators, but they also only make predictions about postseason games (usually the March Madness tournament). Boulier \& Stekler (1999) attempted to predict all March Madness games from 1986 to 1995, excluding the Final Four Championship round (where you might encounter a game where both teams have the same rank). They showed that probit regressions improve the accuracy of the predictions compared to the strategy which consists of simply picking the highest-ranked team over the lowest-ranked team. Caudill \& Godwin (2002) have also worked on predicting winning probabilities from March Madness games (1985 - 1998) excluding the semi-finals and finals games. Their main conclusion was that the skewed logit model with heterogeneous skewness (i.e., the skewness parameter is allowed to vary from one observation to the next) does better than the logit, probit and skewed logit models.
Meanwhile, Caudill (2003) demonstrated that the maximum score estimator does slightly better than the probit model. As for Brown \& Sokol (2010), they used a modified version of the LRMC method to predict the outcome of the 2000 - 2009 March Madness games based on regular season results. We also mention the work of Stekler \& Klein (2012) who innovated by using ``consensus" rankings as predictors, which meant averaging each team's ranking from 29-45 rating systems (depending upon the year). This feature made it unlikely to obtain tied rankings, thus
allowing them to also forecast winners of the final three games of each March Madness tournament.

Schwertman et al. (1996) took a different approach: they used seedings 1-16 in order to predict the probability of each of the 16 seeds winning the regional tournament. They considered 11 different models and evaluated the performance of each. The probability that seed $j$ wins the regional tournament was obtained by adding the estimated probability of each possible ``path" (only one possible opponent in round 1, two possible opponents in round 2, four in round 3 and eight in round 4, for a total of 64 potential paths) with the help of $P(i, j)$ = probability that seed $i$ defeats seed $j$. Along the same lines, Carlin (1996) also focused on predicting regional championship probabilities, but this time based on two important variables: 1) team rankings based on several rating systems like the ``Rating Percentage Index" (RPI) or Sagarin's, rather than regional seeding that can only vary from 1 to 16; 2) spreads (i.e., Las Vegas lines) prior to the first round of the March Madness tournament.

More recently, Lopez \& Matthews (2015) built on the work by Carlin (1996) to beat 400 competing submissions on Kaggle, a free analytics contest, for the 2014 NCAA March Madness tournament. In summary, they merged information from pre-tournament Las Vegas lines along with possession based team efficiency metrics by using logistic regressions. As a matter of fact, over the last years there have been a large number of papers proposing various statistical approaches (from more classical approaches to more machine-learning type of predictions) to predict the outcome of the
March Madness tournament; the interested reader can have a look at the special issue of the ASA's {\it Journal of Quantitative Analysis in Sports}
dedicated specifically to this challenge (Glickman and Sonas, 2015) for a review of various approaches taken to tackle this problem.
Unlike the focus of this paper, the methods proposed in these articles use as many predictors as possible as input, not only the team rankings.

We cite two papers which, unlike the ones above, predict margin of victory. First, Smith \& Schwertman (1999) demonstrated that seed numbers alone provide fairly reliable predictions. Second, Harville (2003) forecasted 93 postseason tournament games (March Madness and NIT tournaments) in 2000 based on more than 4,000 regular season games. The root mean squared error
(RMSE) of the betting line was beaten by basic least squares and by a modified least squares procedure, but they admitted that this phenomenon might be the result of limited sample size (only one year of data).

Our goal is to answer a question that seems to fall in the gap left between the studies cited in the last two paragraphs: can the margin of victory for regular season games, where range of input is 1 to 351 (as opposed to 1-16 when focusing on March Madness games only), can be reasonably well predicted solely from the playing teams' respective rankings and without stringent model assumptions?
We first start by applying the type of models considered by Smith \& Schwertman (1999) and by Harville (2003) to data from regular season games to see if model assumptions are respected and to
see how RMSE values compare to the pure error's RMSE. We then investigate if added flexibility in the response surface improves the predictions, which could be expected with such a wide range for the rankings.

\section{Data description and exploration}

We have put together a database with the margin of victory (MOV, defined as the number of points scored by
the visiting team minus the number of points scored by the home team) of 6,024 regular season games of
NCAA Division I men's basketball played in the 2014-2015 and 2015-2016 seasons
with a true home team for which the RPI rankings were available for both teams. This specific ranking system was chosen for several reasons. First of all, the NCAA has been using it since 1981, in particular as a tool for selecting teams going to the postseason (the March Madness tournament). Second, unlike other well-known ranking systems like the Associated Press and USA Today Coaches that come up only with top 25 teams, the RPI rankings are provided on all 351 teams. Third, our study aimed at using rankings that were really up-to-date. We were able to find historical {\it weekly} rankings on some methods (like the algorithm-based methods such as Massey, Sagarin and Pomeroy), but RPI {\it daily} rankings were easily obtainable directly from the NCAA's official website.

Because we will try to predict MOV using non-parametric methods that are subject to overfitting, we will
train the models on the data from the beginning of the 2014-2015 season up to the first half of the 2015-2016
season (4,518 games) and use the trained model
to predict the MOV of the 1,506 games of the second half of the 2015-2016 season. We also report results
obtained with two other random partitions of the data into a training set of size 4,518 and a validation set of size 1,506.
We refer to the datasets obtained with the original split as training 1 and validation 1 while those
obtained with the random splits will be referred to as training $j$ and validation $j$, $j=2,3$.

\begin{figure}
\begin{center}
\begin{tabular}{cc}
\includegraphics[height=2.5in]{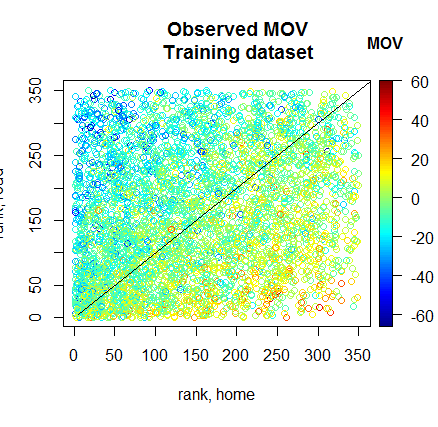} &
\includegraphics[height=2.5in]{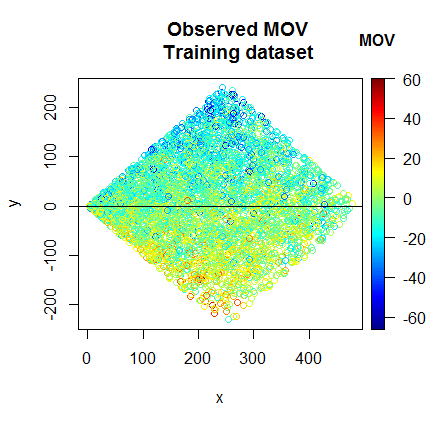}
\end{tabular}

\caption{Observed margins of victory (color) for the training dataset. Left panel: As a function of the rankings $h_{i}$ of
the home team (x-axis) and $r_{i}$ of the visiting team (y-axis), with the line $r_{i}=h_{i}$ drawn in black.
Right panel: Same plot as the left panel, but with a $\pi/4$ counter-clockwise rotation of the axes
and with the line $y_{i}=0$ drawn in black. }\label{fig:obsMOV}
\end{center}
\end{figure}

Let $M_{i}$ denote the MOV for the $i$-th game of a given dataset and $r_{i}$ and $h_{i}$
respectively be the RPI rankings of the road (visiting) and home teams for this game. We suppose that
\begin{equation}\label{eq:modelMOV}
M_{i}=f(r_{i},h_{i})+\epsilon_{i},\ \ i=1,\ldots,n,
\end{equation}
where the $\epsilon_{i}$ are random error terms assumed to be iid from some zero mean distribution.
The observed MOV for the training set are depicted in
the left panel of
Figure \ref{fig:obsMOV},
with the axes representing the
rankings and the color of the symbol representing the margin of victory (blue associated to strongly negative values, which
means a wide margin of victory by the home team, and red associated to strongly positive values, which means a wide margin of
victory by the visiting team). We can see that the color is roughly constant along lines of slope 1, meaning
that the {\it difference} between the rankings of the home and visiting teams contains the bulk of the information.

If we perform a 45 degree counter-clockwise rotation of the axes, i.e., we plot the MOV as a function of
\begin{align}
x_{i}&=r_{i}\sin(\pi/4)+h_{i}\cos(\pi/4)\label{eq:xrot}\\
y_{i}&=r_{i}\cos(\pi/4)-h_{i}\sin(\pi/4)\label{eq:yrot},
\end{align}
we obtain the plot shown on the right panel of Figure
\ref{fig:obsMOV}.

A few key points transpire from Figure \ref{fig:obsMOV}. First and foremost, though the relationship may not be linear
it is clear that there is an association between the MOV and the rankings. Second, smoothing methods will either have to
use very small neighborhoods around the point of prediction or be anisotropic, as we clearly see on the right-hand side panel
that the MOV varies a lot along the $y$-direction but remains fairly constant along the
$x$-direction. Finally, because the MOV surface does not become constant near the edges of the data domain,
smoothing techniques that are less subject to edge effects may potentially lead to better predictions.

\section{Models and predictions}

Table \ref{tab:npres} gives the RMSE obtained when fitting all of the models outlined in this section to each
of the three training sets and each of the three validation sets described above. To put these RMSE into perspective,
the pure error RMSE for each dataset are also provided in the table.

\begin{table}
\begin{center}
\caption{Root mean squared errors on predicted MOV for the various methods
considered in Section 3. Model fitted on the training set $j$ is used to predict
the MOV in the corresponding validation set $j$, $j=1,2,3$.}\label{tab:npres}

\medskip

{\small
\begin{tabular}{lccccccc} \hline
 & Pure error &  Quadratic & Gaussian & Local linear & Isotropic & Anisotropic  \\
Dataset & &  regression & GAM & (LOESS) & kernel & kernel  \\ \hline
Training 1 & 11.52 &	11.51	&	11.49	&	11.49	&	11.56	&	11.54	 \\	
Training 2  & 11.34 &	11.34	&	11.30	&	11.30	&	11.39	&	11.37	 \\	
Training 3 & 11.41 &	11.41	&	11.38	&	11.38	&	11.46	&	11.43	  \\	 \hline
Mean, training  & 11.42 &	11.42	&	11.39	&	11.39	&	11.47	&	11.45        \\ \hline
Validation 1	& 10.92 &	10.96	&	10.92	&	10.95	&	11.00	&	10.97	  \\	
Validation 2  	&	11.47	&	11.50	&	11.49	&	11.50	&	11.47 & 11.45	 \\	
Validation 3 & 11.27 &	11.29	&	11.25	&	11.28	&	11.26	&	11.24	 \\ \hline	
Mean, validation  & 11.22 & 11.25 & 11.22 & 11.24 & 11.25 & 11.22            \\ \hline
\end{tabular}
}
\end{center}
\end{table}

\subsection{The quadratic model}

%

Smith \& Schwertman (1999) and Harville (2003) showed how
a simple quadratic regression model performs very well to predict MOV from team rankings when the range of the rankings
goes from 1 to 16 and when the MOV predicted are those from the March Madness tournament. Our question is whether
this quadratic relation between the rankings and the MOV of regular season games with a true home team
holds on the wider range of rankings from 1 to 351. Fearnhead \& Taylor (2010) did consider a model based on
the full range of rankings, but their goal was to approximate the relationship in order to compute the strength
of the regular season schedule of teams and therefore they settled for a linear regression model.

We consider the quadratic regression model given by
$$
M_{i}=\beta_0+\beta_rr_{i}+\beta_hh_{i}+\beta_{rh}r_{i}h_{i}+\beta_{rr}r_{i}^2
+\beta_{hh}h_{i}^2+\epsilon_{i},\ \ i=1,\ldots,n,
$$
where the $\epsilon_{i}$ are assumed iid $N(0,\sigma^2)$. The interaction in all training sets considered
is not significant and does not improve the training RMSE and is therefore removed.
Thus from hereon it is understood that the quadratic model
does not include an interaction term. When fitted to the original training set (training 1), we obtain
\begin{equation}\label{eq:quad}
\hat{M}_{i}=-5.8
-0.074r_{i}+0.10h_{i}
+4.7\times10^{-5}r_{i}^2-1.2\times10^{-4}h_{i}^2
+\epsilon_{i},\ \ i=1,\ldots,n,
\end{equation}
Its RMSE on the training 1 and validation 1 sets are 11.51 and 10.96, respectively.
To put these figures into perspective, the quadratic models
based on team rankings fitted to Tournament data in different years had an RMSE of 11.12 for Smith \& Schwertman (1999)
and 10.73 for Harville (2003).

This simple model actually turns out to be an excellent summary of the data. Because some pairs of rankings
are repeated, we can proceed to a lack-of-fit test and obtain an estimate of the pure error. The former has
a $p$-value of 0.18 and the estimate of the pure error standard deviation is 11.52 (training 1). It is worth noting, however,
that for the other two train-validation partitions, the $p$-values of the lack-of-fit test were 0.0008 and 0.0445
even if the models' RMSE were virtually equal to the pure error's RMSE.
Smith \& Schwertman (1999) had made a very similar observation, noting that by applying the
calculations of Carlin (1996) to their data, the best RMSE possible was estimated at 11.2.

\begin{figure}
\begin{center}
\begin{tabular}{ccc}
\includegraphics[height=2in]{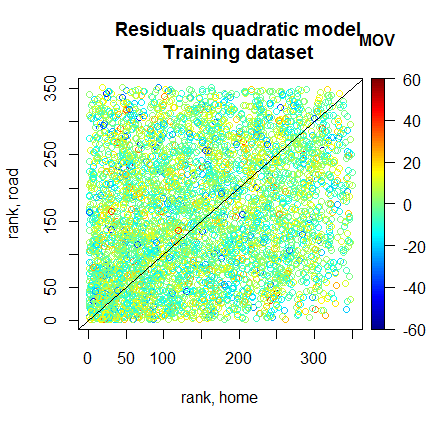} &
\includegraphics[height=2in]{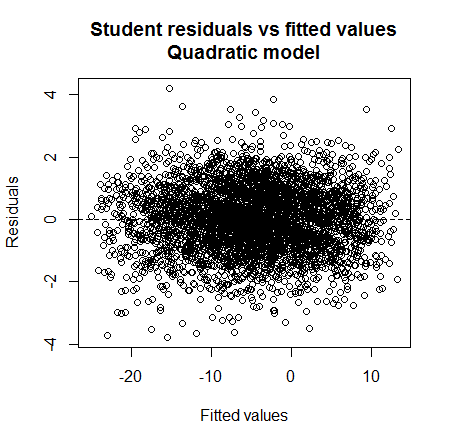} &
\includegraphics[height=2in]{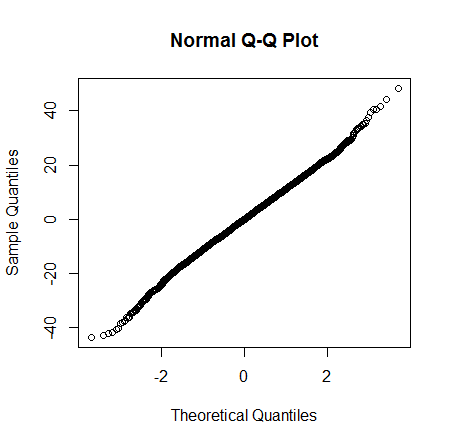}
\end{tabular}

\caption{Residual plots, quadratic model of equation (\ref{eq:quad}) fitted to the training 1 dataset.
Left panel: Residuals as a function of the home and road rankings.
Middle panel: Studentized residuals vs fitted values.
Right panel: Normal quantile-quantile plot of studentized residuals.}\label{fig:resid2}
\end{center}
\end{figure}

The residual plots
depicted in Figure \ref{fig:resid2} do not suggest that there are major problems with the assumption that
the error terms are iid zero mean normal. If we test the normality assumption on the original training set,
the Shapiro-Wilk and Shapiro-Francia test both reject normality with $p$-values$<0.0001$, but the Cram\'er-von Mises test
does not ($p$-value of 0.075). For the second training set, these $p$-values respectively increase to about 0.0003, 0.0002 and 0.204
(and to 0.0004, 0.0002 and 0.078 for training 3).
The previous results suggest that the simple quadratic model captures the essence of the relationship between MOV and the
team rankings, but that slight gains are probably still possible.

In the remainder of this section we explore a few semi- and non-parametric generalizations
of the model to see if we are able to get some reduction in both the training and validation RMSE even though the results
above suggest that this might be a difficult task. For instance Harville (2003) was able to obtain a better RMSE
than that of the quadratic model fitted with ordinary least squares, but when using a modified least squares
approach that included information other than the team rankings. As it turns out, the results of the following sub-sections show
that some reduction in RMSE is still possible, uniformly over all training and validation sets, but these improvements
are minor.

\subsection{Generalized additive model (GAM)}

\begin{figure}
\begin{center}
\begin{tabular}{cc}
\includegraphics[height=2.3in]{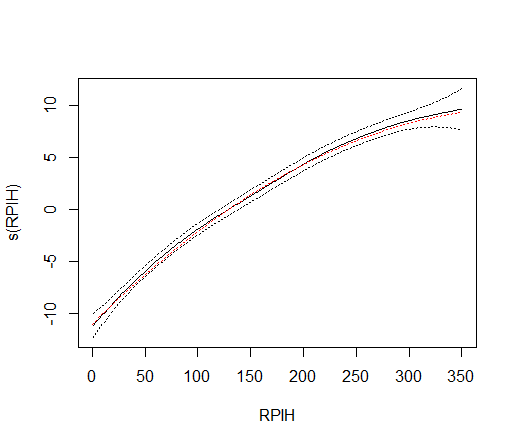} &
\includegraphics[height=2.3in]{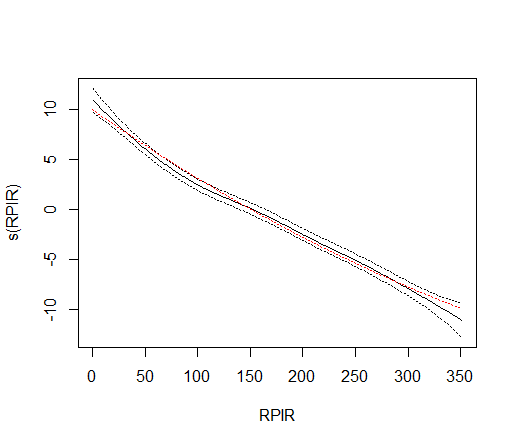}
\end{tabular}

\caption{Part of the mean MOV explained by the
team rankings, GAM model with 95\% pointwise confidence interval
in black and quadratic model in red (training 1). Left panel: rank of home team.
Right panel: Rank of road team.}\label{fig:gamvsquad}
\end{center}
\end{figure}

Because there is no interaction in (\ref{eq:quad}), a natural first generalization of the quadratic model is the
gaussian linear additive model
\begin{equation}\label{eq:gam}
M_{i}=\mu+f_r(r_{i})+f_h(h_{i})+\epsilon_{i},\ \ i=1,\ldots,n,
\end{equation}
where $f_r(\cdot)$ and $f_h(\cdot)$ are arbitrary, but smooth, functions. We fit this model to the training set
by nonparametrically estimating the functions $f_r(\cdot)$ and $f_h(\cdot)$ with smoothing splines using the
backfitting algorithm explained in detail by Hastie (1991, Chapter 7) and implemented in the R package {\tt gam}
(Hastie, 2016).  The added flexibility of this model
results in a decrease in both the original training and validation RMSE, which go down
from 11.51 and 10.96 to 11.49 and 10.92, respectively. Albeit small,
a similar reduction is observed in all training and validation sets considered.
Figure \ref{fig:gamvsquad} compares the estimate of $f_h(h_i)$ obtained with the GAM to that of $\beta_hh_i+\beta_{hh}h_i^2$
obtained with the quadratic model (training 1). While the two models are difficult to distinguish in terms of the effect of the
home team, they differ somewhat more in their estimation of the effect of the road team; in the latter case, the quadratic
model fit almost exits the 95\% pointwise confidence limits provided by the GAM fit.

The exploratory data analyses suggested two possible departures from the model given by (\ref{eq:gam}). First, even though
the interaction was not significant and did not improve the RMSE in the quadratic model, perhaps a more general form
of interaction between the home and road rankings could explain some part of the observed MOV. Second, there seemed to be
an anisotropy whereby the MOV for games involving teams with similar values of $|h_i-r_i|$ are much closer than the MOV
for games involving teams with similar values of $h_i+r_i$. We investigate these extensions to the GAM in the remainder
of this section.

\subsection{Local polynomial smoothing (LOESS)}

\begin{figure}
\begin{center}

\includegraphics[height=2.5in]{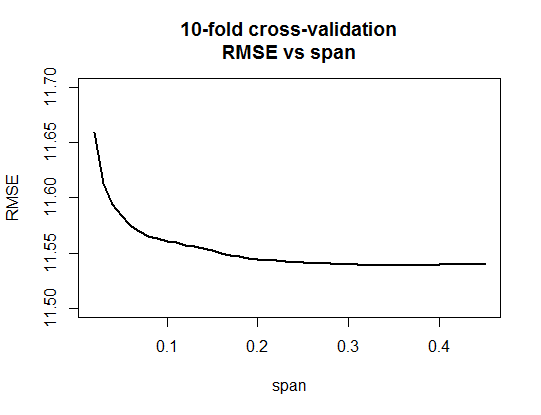}

\caption{RMSE as a function of the span, LOESS method, 10-fold cross-validation
applied to the training 1 dataset.}\label{fig:10foldLOESS}
\end{center}
\end{figure}

We now want to fit a model of the form
\begin{equation}\label{eq:fullbiv}
M_{i}=\mu+f_{rh}(r_{i},h_{i})+\epsilon_{i},\ \ i=1,\ldots,n,
\end{equation}
to our data, without making any parametric assumption on $f_{rh}(\cdot,\cdot)$. There exist several methods
for doing so, but trying all of them would be beyond the scope of our analysis. We first start with the
LOESS method, which consists in fitting local regression models. A detailed treatment of the model and of
the numerical algorithms that can be used to implement the method is given by Cleveland et al (1991, Chapter 8).
For our analysis we use the {\tt loess} function available in R
to fit local polynomials of degree 1 (local linear) because the analyses presented above (e.g., plots
in Figure \ref{fig:gamvsquad}) show that $f_{rh}(r_{i},h_{i})$ does not appear to become constant near the edges of
the data, which could lead to serious bias near the edges if local polynomials of even degree were used. We use
tricubic kernel weights. The smoothing parameter (span) is chosen using 10-fold cross-validation on the training sets.
The plot in Figure \ref{fig:10foldLOESS} shows that the RMSE is not sensitive to the value of the span parameter, as long as it
is between 0.1 and 0.5. We chose a span that keeps 30\% of the data and got a RMSE of 11.49 for the training 1 dataset and 10.95
for the corresponding validation set. In terms of performance, the local linear fit is virtually equivalent to the GAM for all three
training sets. It is close to, but not quite as good as, GAM on the validation sets.

\subsection{Isotropic and anisotropic kernel smoothing}

We now use the general kernel smoothing technique of
Nadaraya-Watson applied to marked point processes over irregular grids (e.g., Diggle, 2013, chapter 5) and implemented in the R package {\tt spatstat}
(Baddeley et al, 2005).
Here the team rankings $\ell_i=(h_i,r_i)$ are viewed as the coordinates of a spatial location of the $i$-th
event of a point process and the MOV $M_i$ is the mark of the process at this location. The prediction (MOV) at a location $\ell_0$
(game involving home team with rank $h_0$ and road team with rank $r_0$) is given by the kernel smoother
\begin{equation}\label{eq:MhatKern}
\hat{M}_0=\frac{\sum_{i=1}^nw_\Sigma(\ell_0,\ell_i)M_i}{\sum_{i=1}^nw_\Sigma(\ell_0,\ell_i)},
\end{equation}
where
\begin{equation}
w_\Sigma(\ell_0,\ell_i)=w^*\left[\left\{{h_0-h_i\choose r_0-r_i}^\top\Sigma^{-1}{h_0-h_i\choose r_0-r_i}\right\}^{1/2}\right]
\label{eq:wSig}
\end{equation}
with $w^*(x)$ the standard normal probability density function.
Thus $\hat{M}_0$ in (\ref{eq:MhatKern}) is a weighted average of all observed MOV, but
with larger weights given to games played with rankings more similar to $(h_0,r_0)$.

The choice of the matrix $\Sigma$ is crucial. Isotropic smoothing gives weights $w_\Sigma(\ell_0,\ell_i)$ that only depend on the
{\it distance} separating $\ell_0$ and $\ell_i$ (and not on the {\it direction} from one point to the other) and amounts to
using $\Sigma=\sigma^2I_2$ with $I_2$ the $2\times2$ identity matrix. To choose the value of
$\sigma$, we apply the leave-one-out cross-validation
algorithm that is already implemented in the package for isotropic smoothing to the training sets.
For the original training-validation partition, we obtained $\sigma=19.4$ and used it with both datasets to obtain RMSE of 11.56 for the training set
and 11.00 for the validation set. This is the worst of all performances seen so far, but this has to be taken with
a grain of salt, as it is not observed in all three splits.

When looking at Figure \ref{fig:obsMOV}, it is clear that the MOV tend to be similar along the south-west to north-east (SWNE) axis
and very different along the north-west to south-east (NWSE) axis. We therefore perform anisotropic smoothing which will make the
weight decrease a lot quicker along the NWSE axis than along the SWNE axis. To do so, we use in (\ref{eq:wSig}) the $(x,y)$
coordinates in the rotated axes as defined by equations (\ref{eq:xrot})-(\ref{eq:yrot}) instead of $(h,r)$ and
$\Sigma={\sigma^2_x\ \ 0\choose 0\ \ \sigma^2_y}$. To find the values of $\sigma_x$ and $\sigma_y$, we applied 10-fold cross-validation
to the training 1 set and obtained $\sigma_x=40$ and $\sigma_y=14$, confirming our intuition that MOV tend to be more similar
along the SWNE axis than along the NWSE axis.

Allowing anisotropy had a relatively minor impact on the RMSE, taking it down
to 11.54 for the training 1 set and 10.97 for the validation 1 set. Even though it is small, this
reduction in RMSE when moving from isotropic to anisotropic smoothing occurs in all six datasets considered.
It is however not good enough to match the performance of the GAM prediction in most cases.
As a matter of fact, we also tried to use the coordinates in the rotated axes $(x_i,y_i)$ in the
quadratic, GAM and LOESS methods instead of the
straight rankings $(h_i,r_i)$. In these cases, the best RMSE were obtained when only using the $y_i$ coordinate, but they were still
not as good as that obtained with each method with the straight rankings $(h_i,r_i)$ and thus are not presented here.

%
%
%
%

\section{Discussion}

The quadratic regression model was already known to be good for modeling the MOV of March Madness tournament games. Our goal in this paper was to investigate whether this was
 also true for regular season games, where the range of the team rankings expands from 1-16 to 1-351. Our analyses confirmed that this is indeed the case.
We observed that the quadratic regression's RMSE matched the pure error's RMSE on all training sets considered and does not lag far behind on the validation sets.

That being said, our analyses also revealed that minor improvements were still possible, and we were able to obtain slight reductions in RMSE with some of the methods
considered. GAM was able to achieve the best RMSE over all three training sets. On validation sets, GAM and anisotropic smoothing performed better than the other procedures tested.
Because of their good overall performance and because they are simple and interpretable, GAM seem to be the most appealing alternative to the quadratic model among the approaches
that were investigated here.

Of course if one starts considering more information than just the team rankings, significant improvements in RMSE can presumably be achieved. For example, the betting line
on our original training and validation sets scores RMSE of 10.54 and 10.34, respectively, and with a huge number of variables (teams' previous records, results of
face-to-face games, injuries to key players, game preview texts, etc.), predictive methods have been able to match or even beat the betting line
over short time periods, such as a
given March Madness tournament.

Other nonparametric prediction methods (e.g., nearest neighbor, thin plate splines, etc.) have been known to perform well in problems of low to moderate dimension and could
certainly have been put to the test in this study. But as we have seen, to predict the MOV of regular season games in NCAA basketball solely from the opposing teams' rankings,
the good old quadratic regression model plays a solid defense and does not leave much room for improvement to its competitors.

\section*{References}

\setlength{\parindent}{0pt}

Baddeley, A., Rubak, E. and Turner, R. (2015). {\it Spatial Point Patterns: Methodology and Applications with R}. London: Chapman and Hall/CRC Press.

Boulier, B. L. and Stekler, H. O. (1999). Are sports seedings good predictors?: an evaluation. {\it International Journal of Forecasting}, {\bf 15}, 83-91.

Brown, M. and Sokol, J. (2010). An Improved LRMC Method for NCAA Basketball Prediction. {\it Journal of Quantitative Analysis in Sports}, {\bf 6}, Article 4.

Carlin, B. P. (1996). Improved NCAA Basketball Tournament Modeling via Point Spread and Team Strength Information. {\it The American Statistician}, {\bf 50}, 39-43.

Caudill, S. B. (2003). Predicting discrete outcomes with the maximum score estimator: The case of the NCAA men's basketball tournament. {\it International Journal of Forecasting}, {\bf 19}, 313-317.

Caudill, S. B. and Godwin, N. H. (2002). Heterogeneous skewness in binary choice models: Predicting outcomes in the men's NCAA basketball tournament. {\it Journal of Applied Statistics}, {\bf 29}, 991-1001.

Cleveland, W. S., Grosse, E. and Shyu, W. M. (1991). Local regression models. Chapter 8 of {\it Statistical Models in S}, eds J.M. Chambers and T.J. Hastie, Wadsworth \& Brooks/Cole.

{d}el Corral, J. and Prieto-Rodriguez, J. (2010). Are differences in ranks good predictors for Grand Slam tennis matches? {\it International Journal of Forecasting}, 26, 551-563.

Diggle, P. J. (2013). {\it Statistical Analysis of Spatial and Spatio-Temporal Point Patterns, 3rd ed}. Chapmal and Hall/CRC, Boca Raton.

Fearnhead, P. and Taylor, B. M. (2010). Calculating strength of Schedule, and choosing teams for {M}arch {M}adness.
{\it The American Statistician}, {\bf 64}, 108-115.

Glickman, M. E. and Sonas, J. (2015). Introduction to the NCAA men's basketball prediction methods issue. {\it Journal of Quantitative Analysis
in Sports}, {\bf 11}, 1-3.

Harville, D. A. (2003). The Selection or Seeding of College Basketball or Football Teams for Postseason Competition. {\it Journal of the American Statistical Association}, {\bf 98}, 17-27.

Hastie, T. J. (1991) Generalized additive models. Chapter 7 of {\it Statistical Models in S}, eds J. M. Chambers and T. J. Hastie, Wadsworth \& Brooks/Cole.

Hastie, T. J. (2016) {it gam: Generalized Additive Models}. R package version 1.14. https://CRAN.R-project.org/package=gam.

Lopez, M. J. and Matthews, G. J. (2015). Building an NCAA men's basketball predictive model and quantifying its success. {\it Journal of Quantitative Analysis in Sports}, {\bf 11}, 5-12.

Schwertman, N. C., Schenk, K. L. and Holbrook, B. C. (1996). More Probability Models for the NCAA Basketball Tournaments. {\it The American Statistician}, {\bf 50}, 34-38.

Smith, T. and Schwertman, N. C. (1999). Can the NCAA Basketball Tournament Seeding be Used to Predict Margin of Victory? {\it The American Statistician}, {\bf 53}, 94-98.

Stekler, H. O. and Klein, A. (2012). Predicting the Outcomes of NCAA Basketball Championship Games. {\it Journal of Quantitative Analysis in Sports}, {\bf 8}, 1-10.

\end{document}